# Crystalline hydrogenation of graphene by scanning tunneling microscope tip-induced field dissociation of $H_2$

S. J. Tjung[1], S. M. Hollen[2], G. A. Gambrel[1], N. M. Santagata[1], E. Johnston-Halperin[1], J.A. Gupta[1]*

[1]Department of Physics, The Ohio State University, Columbus, OH 43210

[2]Department of Physics, University of New Hampshire, Durham, NH 03824

Abstract

We have developed a novel method for crystalline hydrogenation of graphene on the nanoscale. Molecular hydrogen was physisorbed at 5 K onto pristine graphene islands grown on Cu(111) in ultrahigh vacuum. Field emission local to the tip of a scanning tunneling microscope dissociates $H_2$ and results in hydrogenated graphene. At lower coverage, isolated point defects are found on the graphene and are attributed to chemisorbed H on top and bottom surfaces. Repeated $H_2$ exposure and field emission yielded patches and then complete coverage of a crystalline √3 x √3 R30° phase, as well as less densely packed 3x3 and 4x4 structures. The hydrogenation can be reversed by imaging with higher bias voltage.

*Corresponding author: Tel: 614 247-8457. E-mail: jgupta@physics.osu.edu

# 1. Introduction

The interaction of hydrogen and pure carbon in amorphous or crystalline allotrope forms has been studied in a variety of contexts including interstellar catalysis [1], hydrogen storage [2], quantum fluids [3, 4] and tunable nanomaterials [5, 6]. Early studies of $H_2$ on graphite have explored a variety of gas/liquid/solid phase transitions at low temperature (< 20K) and low coverage (< 1 monolayer) [7, 8], and similar behavior has been predicted for $H_2$/graphene [9]. The interaction with atomic hydrogen is also of considerable interest for tuning graphene's electronic, magnetic, and chemical properties for a variety of applications.[6, 10–13] Crystalline ordering would help in identifying the intrinsic properties of hydrogenated graphene (H-Gr) and mitigating effects of disorder on, for example, carrier mobility in electronic transport. Currently, the evidence for ordered H-Gr (or graphane at complete coverage) is somewhat limited. Most reports of H-Gr via electron beam irradiation [14–16], chemical [17], plasma,[5, 18–20] or thermal cracker [21–33] methods indicate insulating behavior [5, 16, 20, 21, 24, 25, 27] but little or no evidence for crystalline order. STM experiments have shown some ordering associated with Moiré patterns[27] or in ~$nm^2$ patches,[32] and other techniques have provided indirect evidence of ordered structures.[26, 34]

Here we report a new, *in situ* method for hydrogenation of graphene that results in crystalline order. Atomic hydrogen was produced by dissociating physisorbed $H_2$ with field emission local to the tip of a scanning tunneling microscope (STM) at low temperature (5 K). This process generates point defects on graphene with two apparent heights, which we attribute to chemisorbed H on top and bottom surfaces. With higher hydrogen coverage, we observed a √3 x √3 R30° surface which we attribute to a lattice of hydrogen dimers, as well as 3x3 and 4x4 lattices corresponding to less dense structures. We found that pristine graphene could be recovered by imaging at higher bias voltages, suggesting that this hydrogenation method may be useful for locally and reversibly patterning graphene's surface properties.

## 2. Experimental Details

Graphene islands were grown on Cu(111) by thermal decomposition of ethylene in ultrahigh vacuum (UHV) at a base pressure of ~$10^{-10}$ mbar.[35–37] This method produces pristine graphene and interfaces which can be studied by STM without any exposure to air.[35, 36] A clean Cu(111) surface was first prepared by cycles of $Ar^+$ sputtering and annealing at 600 °C, repeated until surface contamination was minimized in Auger electron spectra. Graphene was grown by introducing $2 \times 10^{-5}$ mbar of ethylene gas into the UHV chamber while cycling the sample temperature 2-4 times from room temperature to ~950 °C, resulting in an average coverage < 0.25 monolayers. The sample was then precooled to ~ 120 K within 30 minutes and transferred into the cold (5 K) STM, where the effective pressure is much lower than the gauge pressure of $10^{-10}$ mbar. All measurements were made with a CreaTec LT-STM operating at 5 K. Constant current STM images were collected with cut PtIr or etched W tips. STM images were calibrated using atomically resolved images of the Cu(111) surface. In our experience, atomic resolution imaging of graphene/Cu(111) under typical tunneling conditions likely results from molecule-terminated tips, and we distinguish these from metal atom-terminated tips by imaging and spectroscopy on the Cu surface.[35, 36] Image analysis was performed with the WSxM software[38].

A precision leak valve admitted $H_2$ into the STM chamber at a pressure of $10^{-7}$ mbar for 5-10 minutes. A wobble stick is used to open small holes in the radiation shields surrounding the STM that allow some of the $H_2$ to adsorb onto the sample surface at 5 K.[39] Dissociative adsorption of $H_2$ has a significant activation barrier on both Cu (~0.4 eV)[40, 41] and graphene (~0.2 eV) [42, 43], so that under these conditions, $H_2$ is only weakly bound to the surface

(physisorption). STM images discussed below suggest a coverage of ~ 1 monolayer, but the mobility of physisorbed $H_2$ on the surface makes a precise estimate difficult because individual molecules cannot be imaged with STM. In contrast to previous efforts at electron-induced hydrogenation [14–16], here we produce atomic hydrogen by *in situ* field-emission dissociation of $H_2$ by the STM tip. The STM has previously been used to extract H from hydrocarbons [44] and desorb H from passivated semiconductor surfaces [45, 46]. Dissociation via electronic excitation and electron attachment have also been studied theoretically, and tabulated cross sections for $H_2$ indicate maxima in the few eV and ~ 30-70 eV ranges depending on the process.[47] Electric field strengths of the order used here (~ V/nm) have also been predicted to catalyze hydrogenation of graphene by lowering the chemisorption activation barrier.[48] In our experiments, we dissociate $H_2$ by operating the STM in the field emission regime: the tip is retracted ~130 nm from the surface and a relatively high voltage of 50-120 V is applied to achieve a field emission current of ~0.5 nA for 5-10 minutes. The same atomic scale area can be imaged before and after field emission to unambiguously determine surface changes. Both cut and etched tips were used to qualitatively study how the tip shape influences the field dissociation process, and more details and control experiments are discussed in the Supporting Information.

### 3. Results and discussion

To demonstrate production of atomic H via field dissociation, we first discuss control experiments on Cu(100), where chemisorbed atomic H was well characterized with STM.[44] An STM image of the clean Cu(100) surface is shown in Figure 1a. Point defects in the image are mostly residual CO from the UHV chamber, and atomic resolution of the Cu surface was not obtained, as expected for these imaging conditions. Figure 1b is the same area after exposure to $H_2$. The rapid motion of physisorbed $H_2$ even at 5 K prevents imaging of isolated molecules with

STM, and makes it difficult to directly measure the coverage. However, a stable 3.6 Å triangular lattice is observed which is similar to the intermolecular spacing in solid $H_2$, and was attributed to approximately monolayer coverage in previous studies.[39] Field emission creates new point defects that are visible as shallow depressions in Fig. 1c, which we attribute to chemisorbed atomic H. Though stable enough for STM imaging, we found atomic H hops on the surface at a rate consistent with quantum tunneling as established previously.[44] We verified that field emission

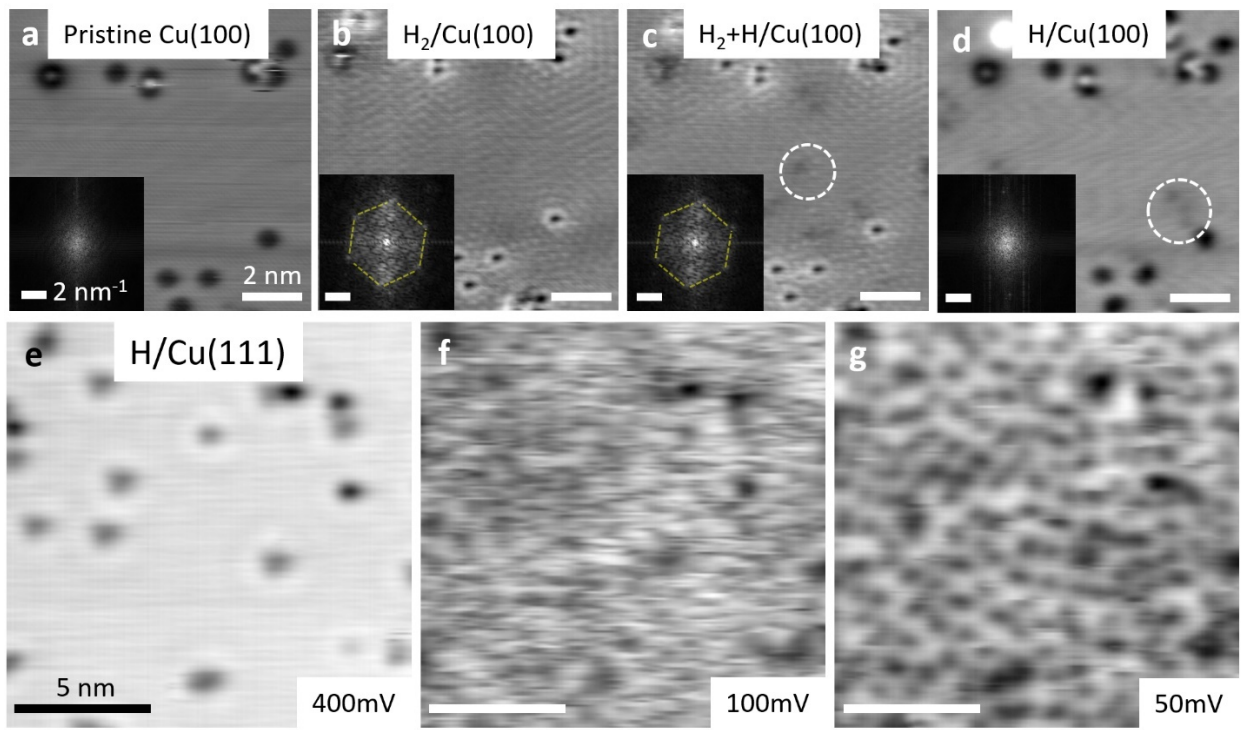

**Figure 1. Field dissociation of $H_2$ on Cu** (a-d) STM images and corresponding FFTs of Cu(100) at each step of the process. (20 mV, 0.2 nA) (a) pristine surface showing residual chamber contaminants as markers (b) 3.6 Å triangular lattice following $H_2$ adsorption (c) generation of chemisorbed atomic H after field emission. The dotted circle indicates one H atom. The $H_2$ lattice is still observed in the real space and FFT images. (d) after desorption of physisorbed $H_2$ by warming up to 20K. The $H_2$ lattice is no longer observed. The dotted circle indicates three H atoms remaining on the surface. (e-g) STM images of Cu(111) after a complete field dissociation cycle ($H_2$ deposition, field emission, $H_2$ desorption). (e) At 400mV bias, H atoms are strongly perturbed by the tip and do not appear in the image. (f) At 100 mV bias, chemisorbed H are more apparent, but still perturbed by the tip during scanning. (g) At 50mV bias, H atoms can be stably imaged as a high density of dark depressions. (0.2nA)

on the clean Cu surface does not produce the atomic hydrogen shown in Fig. 1. The field emission process is local to the ~ micron-scale area near the tip, and we only find atomic H on the surface near the tip's location. Figure 1c also indicates that the $H_2$ lattice remains, which we attribute to the fact that $H_2$ is mobile and can rapidly diffuse to the tip's location, replenishing the local supply as $H_2$ is dissociated. To desorb excess $H_2$ on the surface, we can warm up the STM stage to 20 K.[39] Subsequent STM imaging at 5 K (Fig. 1d) shows that the atomic H remains chemisorbed on the surface, but there is no evidence for the physisorbed $H_2$ layer in STM images or spectroscopy.[39]

We next demonstrate that atomic H can be generated in a similar manner on the Cu(111) substrate to be used for graphene growth. In contrast to Cu(100), atomic H on Cu(111) is more difficult to identify because it hops more quickly and is easily perturbed by the STM tip during imaging.[41] Figure 1e shows an STM image of a Cu(111) region following the cycle of $H_2$ adsorption, field dissociation and warming to 20 K. Under these imaging conditions (i.e., 400 mV bias), individual H are not resolved, and the STM image only shows residual surface contaminants and what otherwise would appear to be clean Cu(111). Lowering the bias voltage to 100 mV however, results in streaky imaging suggestive of tip-induced adsorbate motion (Fig. 1f). This is confirmed by further lowering the bias voltage (e.g., 50 mV in Fig. 1g), where the chemisorbed H atoms can now be stably imaged as a high density of shallow dark depressions. This bias-dependent behavior is not observed on pristine Cu(111), and is consistent with recent studies of H on Cu(111) produced by a different method. [41]

Turning now to the interaction of H/H$_2$ with graphene, Figures 2-5 discuss a sequence of experiments on a ~ 3000 nm$^2$ graphene island, subject to three of the field dissociation cycles illustrated above. As a baseline for comparison, an STM image of the pristine graphene island is shown in Fig 2a. Graphene islands are readily distinguished from the Cu(111) surface by the greatly reduced density of point defects,[35],[36] and this is confirmed by atomically resolved images, which show contrast at the honeycomb centers spaced by the expected 2.46 Å (c.f. Supporting Information Fig. S1), and a Moiré pattern consistent with a rotation of 0° with respect to the underlying Cu(111) lattice (c.f., Fig. S2). STM images of graphene at low voltage are dominated by standing wave patterns associated with the scattering of the underlying Cu(111) surface state electrons, which produces a ring in the corresponding FFT [35]. Spherical scattering was used to assign faint depressions appearing on the graphene islands to defects remaining in the underlying Cu [36]. Surface state scattering is not as clear for the larger depressions which we attribute to defect clusters in the graphene. Though not apparent under the conditions in Fig. 2a, STM images of this area at higher bias (c.f., Fig. S2a) also show bright point defects which we attribute to carbon vacancies formed during growth.[36] The positions of these vacancies are marked with arrows in Figure 2a to guide the eye.

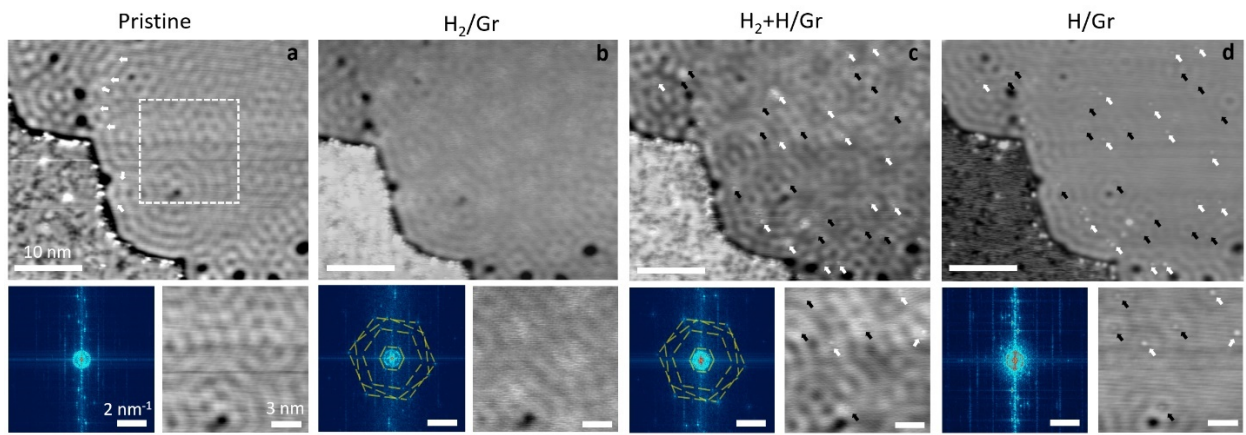

**Figure 2. Field dissociation cycle on Gr/Cu(111)** (a) Pristine graphene island. Bottom right image is a magnified view of the area indicated in the white square. Bottom left image is the FFT showing a circular ring from scattering of the Cu(111) surface state electrons. Arrows show vacancy positions inferred from other imaging conditions (b) Following H$_2$ deposition at 5K. The magnified view shows indications of H$_2$ lattices, evident as multiple hexagons in the FFT. (c) Chemisorbed H following field emission. Black (white) arrows indicate chemisorbed H atoms with dim (bright) image contrast. The FFT still shows H$_2$-related lattices as in (b). (d) Desorption of H$_2$ at 20K. H$_2$ lattices are no longer observed in real space or FFT images, leaving chemisorbed H behind. (0.2nA, 50mV)

Figure 2b shows the same graphene island after adsorption of $H_2$ at 5 K. As on Cu, physisorbed $H_2$ appears to be mobile on graphene, and isolated molecules are not resolved in STM images. FFT analysis reveals a hexagon associated with the 3.6 Å lattice similar to Cu, but also additional hexagons corresponding to triangular lattices with 4.0 Å, 5.2 Å, and 10 Å periodicities that were not observed on Cu. These hexagons have distinct rotational orientations: 49°, 4° and 11° respectively, while the 3.6 Å lattice is rotated by 5° with respect to the graphene. The distinct rotation angles and lattice spacings suggest each physisorbed structure is incommensurate with the others. Inverse FFT images do not show any clear domain structure, and all periodicities extend over the entire graphene region. These additional lattices could be due to multiple layers of condensed $H_2$ in the incommensurate solid region of $H_2$/graphene phase diagram.[7] The FFT also shows spots near the center associated with the graphene Moiré pattern, which is unaffected by $H_2$ adsorption (c.f., Fig. S2).

Field emission on the $H_2$/Graphene surface generates new point defects with bright image contrast, as indicated by arrows in Fig. 2c. Physisorbed $H_2$ remains on the surface, and the FFT still indicates the multiple triangular lattices. To better study the new point defects, $H_2$ was desorbed by warming to 20K, and the result is shown in Fig. 2d. We counted ~ 64 new point defects on this island, which can be sorted into two categories by their apparent height (0.15 ± 0.05 Å and 0.24 ± 0.05 Å) (c.f., Fig S3), and occur in nearly equal numbers (31 and 33 respectively).

The new defects are stable under typical imaging conditions and did not exhibit any motion in the several days we studied this region.

Figure 3 shows an atomic resolution STM image of five of these new defects in this same graphene island. The centers of the honeycomb lattice are imaged with dark contrast in this image, and we use this to locate an overlay of the graphene lattice, with an origin on the right of the image as indicated. There is very good registry of the overlay on the right side of the image, indicating uniform graphene and low drift during imaging. Both types of the defects are imaged with a similar three-fold symmetric contrast, and two distinct orientations, rotated by 180°. Though the details of the contrast are likely affected by the tip termination, the symmetry is characteristic of carbon-site point defects [18, 49], and in fact the overlay indicates that the three defects near the origin point appear centered on carbon sites. The orientation of the defect contrast is correlated with the sublattice site, as indicated by the triangles in the figure. The graphene lattice appears to be slightly distorted in the vicinity of such defects, which shows up as a reduced registry of point defects away from the origin point in our overlay, as well as a phase slip in registry of the graphene lattice on the other side of the image.

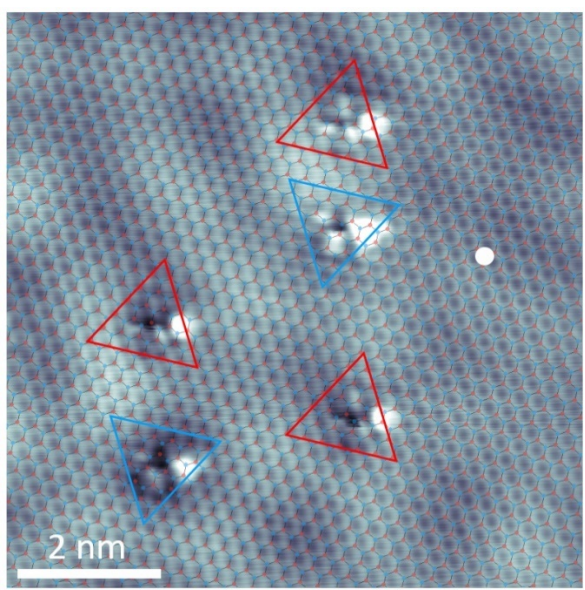

Figure 3. Atomic resolution image of chemisorbed H and graphene. An overlay is registered to the image at the origin point indicated by the white dot. A slight distortion in the graphene lattice due to the defects causes errors in the registry on the left side of the image. Red and blue triangles indicate H on different sublattice sites.

We attribute these defects to chemisorbed atomic hydrogen, which has both a substantial activation barrier (~ 0.2 eV) and adsorption energy (~ 0.7 eV) atop the carbon site [42, 43, 50]. Typically, $H_2$ is dissociated on a hot (> 2000 °C) filament into atomic H, which then has enough kinetic energy to overcome the chemisorption barrier. Previous STM images of chemisorbed H on graphene exhibit three-fold symmetry and two distinct orientations depending on the sublattice site [18, 51], similar to our observations. In our experiments, field dissociation of $H_2$ likely produces atomic hydrogen with sufficient energy to overcome the activation barrier and chemisorb on graphene. This could happen due to inelastic scattering processes during the electronic excitation [47]. We suggest that the two distinct apparent heights correspond to hydrogen chemisorbed on the top or bottom surfaces of the graphene. Graphene is thought to lie ~ 3.25 Å above the Cu(111) surface, which leaves room for the ~ 1.1 Å C-H bond expected for chemisorbed hydrogen.[6] We attribute the fainter defects to chemisorbed H in the interfacial region, and the brighter defects to chemisorbed H on the top graphene surface, pointing toward the tip.

In addition to isolated point defects, we observed crystalline hydrogen structures that are commensurate with the graphene lattice. Figure 4 shows the predominant structure in a sequence of images taken of this same graphene island after additional exposure to $H_2$ and a second and third round of field emission to create more atomic hydrogen on the surface. Figure 4a shows a corner of the island after the second round of field emission; no additional point defects were created in this area, and the 'scratchiness' of the image likely reflects perturbation of adsorbed hydrogen by the STM tip during scanning. In repeated imaging, dark patches appeared near the island edge, which appear ~ 0.2 Å lower than the surrounding graphene (Fig. 4b). The dotted arrow in Fig. 4b shows the trajectory of the patch in a sequence of images. We observed little if any patch motion between consecutive images separated by 10 hrs, suggesting the trajectory is due to the STM tip's influence during scanning rather than intrinsic diffusion processes. An atomic resolution image of the patch within this sequence is shown in Fig. 4c with the corresponding FFT in Fig. 4d. A triangular lattice with a spacing of 4.3 Å is observed that is consistent with a √3 x √3 R30° commensurate structure on graphene (√3·$d_o$ = 4.26 Å, where $d_o$=2.46 Å is the graphene lattice constant). Several bright point defects in the patch could be associated with vacancies, and we confirmed these were not present in the graphene. Perturbation by the tip continues to push the crystalline patch toward the graphene edge (Fig. 4e), until it is no longer

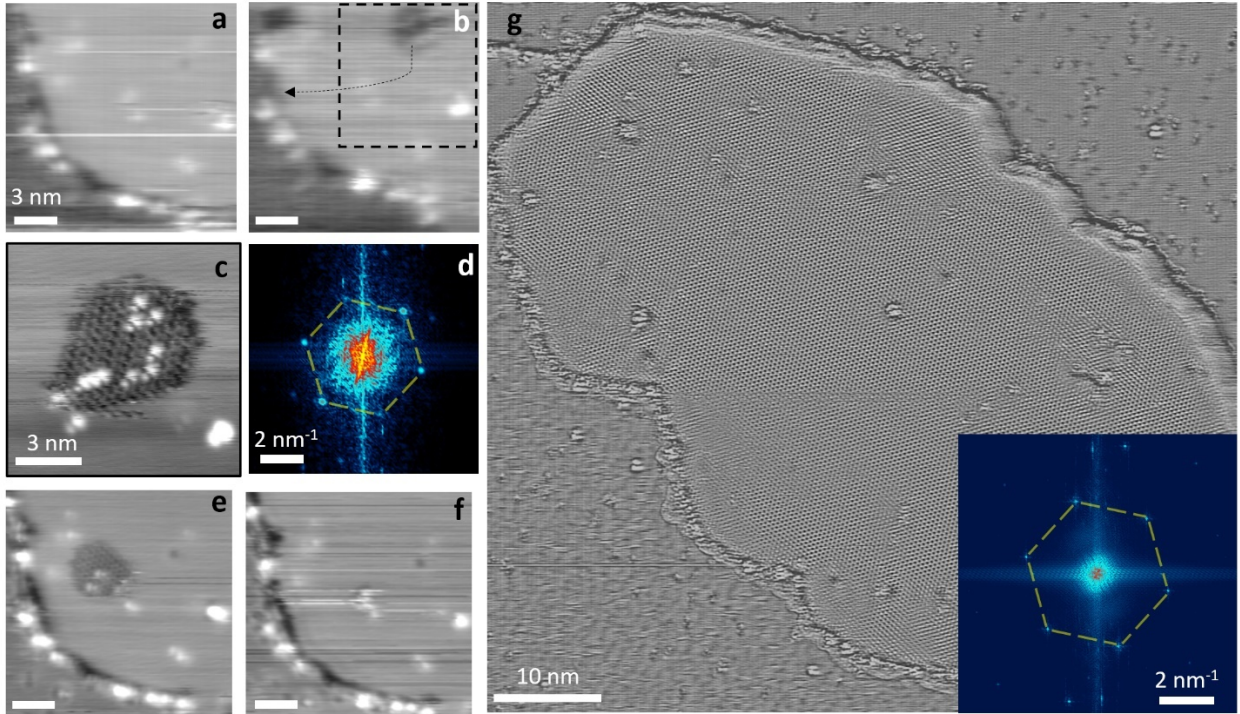

Figure 4. Crystalline hydrogenation of graphene after additional $H_2$ exposure and field dissociation. (a)-(f) Sequential imaging of the same area showing formation of an ordered H-Gr patch by tip perturbation of H on the surface. (1V, 0.2nA) (a) Image showing increased 'streakiness' due to perturbation of hydrogen on the surface during scanning. (b) Image showing dark patches near the top left and top right corners. The arrow indicates the patch trajectory in subsequent imaging. (c) higher magnification atomic resolution image of the patch, in the region marked by the box in (b). The patch has grown larger, and shows a triangular 4.3Å lattice (d) corresponding FFT of (c) showing a new hexagonal spot pattern. (e) Image showing the patch has now moved closer to the island edge. (f) Image showing the patch is no longer clearly resolved. (g) Laplace filtered STM image of the graphene island after a third round of field dissociation. The 4.3Å lattice is present over the entire island. (50mV, 0.2nA)

clearly observed (Fig. 4f). Following this sequence of images, a third round of field emission was conducted on this island, with the result that a stable 4.3 Å lattice was observed over the entire island (Fig. 4g). There are a few new defects created, but there is no clear indication of the isolated chemisorbed H from the preceding rounds of treatment. The Moiré pattern remains visible in both the real-space and FFT images, and was unchanged from the pristine graphene (c.f., Fig. S2). Though the structure was stable for several days of imaging, it could be perturbed by the STM tip under some imaging conditions to reveal the underlying graphene (c.f., Fig. S4).

To gain some insight into the lattice structure, Figure 5a shows an STM image with simultaneous atomic resolution of a boundary between an H-Gr patch and graphene, evident as two hexagons in the corresponding FFT (Fig. 5b). To more precisely determine the atomic structure, we inverse FFT-filtered the image on both Gr and H-Gr spatial frequencies, resulting in Fig. 5c. We assign bright contrast on graphene to the honeycomb centers, and extend an overlay onto the hydrogenated graphene region. Though we cannot completely rule out artifacts due to the tip termination, bright contrast appears centered between carbon sites, which we discuss below could indicate that hydrogen is adsorbed as dimers in the √3 structure.

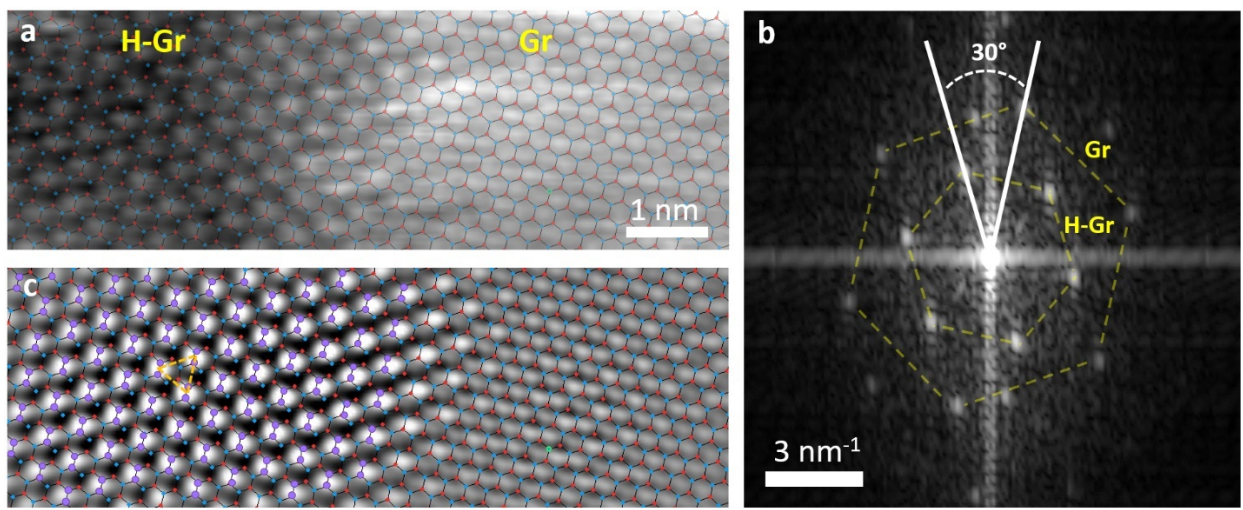

Figure 5. Atomic resolution image of a H-Gr patch edge. (a) raw image indicating H-Gr and Gr regions. (b) corresponding FFT showing spot patterns for both lattices (c) inverse FFT-filtered image selected to include both sets of lattice spots to better register an overlay lattice to the graphene. This is then extended onto the H-Gr region to attribute bright contrast with a √3 lattice of bridge sites. The same overlay is added to (a) for reference. (500mV, 0.2nA)

Figure 6 shows two other crystalline structures that were occasionally observed. The pristine graphene area in Fig. 6a was subjected to one complete cycle of $H_2$ adsorption, field dissociation and 20 K warm-up. Afterward, Figure 6b showed two regions with triangular lattice spacings of 7.8 Å and 10.1Å, close to 3x3 and 4x4 reconstructions of graphene. In both domains, the Moiré pattern seen in the pristine graphene is no longer observed, in contrast to the √3 structure (Fig. 4g). Similar to a previous study [21], we find that the graphene hydrogenation is reversed by imaging with higher bias voltage (>4.5 V) compared to typical imaging conditions (<1 V). As we repeatedly imaged the area in Figure 6 with higher bias (not shown), the Moiré pattern from the pristine graphene gradually re-emerged. After switching back to typical imaging conditions (Fig. 6c), the 3x3 and 4x4 reconstructions were no longer observed, and the pristine graphene was recovered, with a few additional point defects (Fig. 6c). The reconstructions in Figure 6b correspond to a less dense packing of the hydrogen, suggesting that factors such as $H_2$ coverage,

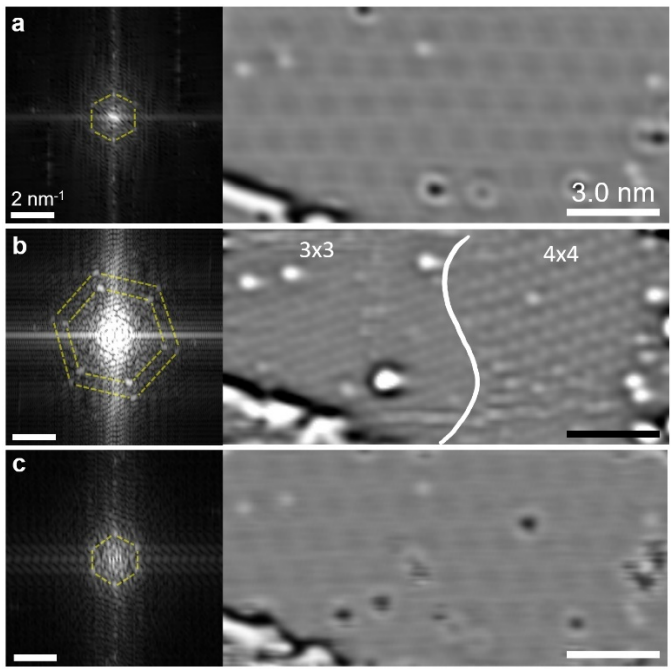

Figure 6. Reversibility of graphene hydrogenation and other structures. (a) Laplace filtered STM image of a pristine graphene, different from the one in Figs. 2-5. (inset) FFT showing hexagonal spots corresponding to the Moiré pattern (b) Image after the hydrogenation process, showing two regions with triangular lattice spacings corresponding to 3x3 and 4x4 reconstructions. (Inset) FFT showing corresponding hexagonal spot patterns (c) After imaging with high bias voltage (> 4.5V), the 3x3 and 4x4 reconstructions are no longer observed and the pristine graphene is recovered. (Inset) FFT showing the same Moiré pattern as in (a). (1V, 0.2nA)

temperature or field emission current density local to the STM tip may influence the hydrogenation process.

To interpret these crystalline structures, we note that a variety of structures have been predicted, depending on the hydrogen coverage, and whether one considers single- or double-sided adsorption [52]. Clustering of chemisorbed atomic hydrogen on graphene is strongly preferred by ~ 1 eV / atom, provided adsorption occurs on the other sublattice (i.e., *ortho-* and *para-*sites in reference to the benzene ring) [50, 53]. *Lin et al.* recently reported STM images of small (~1 nm$^2$) patches of a √3 structure on Gr/Cu foils that were attributed to *ortho-*dimers, with the hydrogen atoms on opposite sides of the graphene sheet, as well as two 1x1-like structures [32]. In contrast, we predominantly observed much larger areas of the √3 structure (e.g. >3000 nm$^2$ in Fig. 4g), and occasionally less densely packed areas (Fig. 6). One possible explanation for our distinct observations is that the thermodynamic conditions in the two methods are quite different. The typical thermal cracking method used by *Lin et al.,* results in a flux of highly energetic, reactive H onto the surface, which was held at room temperature. Here, we produce H *in situ* from H$_2$ gas already physisorbed on the surface at 5 K, and electron scattering / electric field effects dictate the adsorption energetics near the tip. The pristine graphene/Cu interface in our studies is also distinct from prior efforts, and substrate influences on hydrogen adsorption have yet to be considered in detail experimentally [27, 34] or theoretically. The formation of double-sided *ortho-*dimers is consistent with the lattice registry in Fig. 5, and our observation that the field dissociation method equally generates top- and bottom chemisorbed H (e.g., Fig. 3). We expect that due to their higher binding energy, *ortho-*dimers would preferentially form with the higher H coverage produced by multiple rounds of our field dissociation process as illustrated in Fig. 4. The patch formation and motion shown in Figure 4 is unexpected in light of the sizeable activation (~ 0.2eV) and diffusion barriers (~ 1eV) [54] for chemisorbed H , but these barriers may be lowered in the vicinity of other hydrogen [50], and may be further influenced by the STM tip, where electric fields of the order present in our experiments (~ V/nm) have been predicted to influence hydrogen adsorption [48, 55].

Surprisingly, we observed no significant changes in the graphene lattice constant associated with the √3 structure. The measured spacing is within experimental error of √3$d_0$ for pristine graphene, and there are no apparent changes in the graphene islands' area or Moire pattern when it is fully hydrogenated. Though the interaction of graphene with the Cu substrate is relatively weak [35], we speculate that there could be enough residual strain to counteract hydrogen-induced changes in the lattice constant. We also note that we cannot completely rule out the influence of physisorbed $H_2$, which may remain on the surface at sub-monolayer coverage despite our annealing procedure. Though a commensurate √3 structure exists in the predicted $H_2$/graphene phase diagram [9], we did not observe this structure when only adsorbing $H_2$ onto the surface (Fig. 2b), and the observed registry (Fig. 5) does not agree with the expected hollow adsorption site for $H_2$/graphene [9]. Because these phases are coverage-dependent [56], we also performed a control run where we imaged the Gr/Cu surface continuously while slowly adsorbing $H_2$ onto the surface over the course of a full day. While eventually STM images revealed the formation of the 3.6 Å lattice, the √3 structure was never observed. Thus we conclude that the √3 structure, which only occurs after field dissociation, must be associated with atomic H.

In contrast to previous hydrogenation efforts, our field dissociation method produces crystalline ordering local to the micron-scale field emission profile of the STM tip. Local and reversible hydrogenation of graphene is a promising avenue for patterning graphene's electronic, chemical and magnetic properties. It will be interesting to extend this method to graphene on other substrates to probe the substrate's influence on hydrogen adsorption and the resultant changes in these properties.

**Acknowledgements**

Funding for this research was provided by the Center for Emergent Materials at the Ohio State University: an NSF MRSEC under award number DMR-1420451.

# Crystalline hydrogenation of graphene by STM tip-induced field dissociation of H$_2$
SUPPLEMENTARY MATERIAL

**Repeatability of field emission dissociation**

While the field dissociation process was reproduced many times, the threshold conditions for the process could not be well determined because there are several factors in our experiments which limited the specifics of the hydrogenation process:

1. *The initial coverage of physisorbed H$_2$ is difficult to estimate.* While H$_2$ is admitted into the UHV chamber at a desired gauge pressure via a precision leak valve, only a small fraction of the gas gets to the surface, as it must pass through ~ 1cm diameter holes in the radiation shields which surround the STM. These shields are cooled to 77K and 4K by the lN$_2$ and lHe cryogen reservoirs, and can act as cryopumps for H$_2$. The STM stage itself at 5K can also act as a cryopump, and the capacity/pumping speed of these cryopumps depends on the chamber history, such as time spent cold. The STM tip itself can also shadow direct line of sight to the nanoscale area being imaged, and the physical shape of the tip is not well controlled or characterized *in situ*. Because of these factors, the observed H$_2$ coverage on the surface of the sample can vary for nominally the same dosing parameters (e.g., 10$^{-7}$ mbar of H$_2$ for 5-10 minutes). We also don't have a direct measure of the H$_2$ coverage; the onset of the 3.6 Å lattice is suggestive of monolayer coverage [1], but we were unable to directly image isolated H$_2$ molecules with STM.

2. *Thermal/electron-induced changes to the local H$_2$ coverage are also difficult to quantify.* During the field emission dissociation, we observe that the STM stage temperature rises slightly to 6K measured with a diode thermometer on the stage. We have no way to measure the local temperature at the tunneling junction between the STM tip and the sample, which could be a lot higher than the measured STM stage temperature. Physisorbed H$_2$ starts desorbing around 10K, so the local coverage may be lower due to this heating effect. Electron-induced desorption can also occur, [2] which may affect the local H$_2$ coverage during the field dissociation process.

3. *The field emission dissociation has a strong dependence the nanoscale profile of the STM tip.* While the STM tunnel junction will always reflect the most protruding atom(s) on the tip, sharp asperities further up on the tip may dominate the field emission. In one experiment with a cut PtIr tip, we observed a partially hydrogenated graphene island located ~1 μm away from scan area. This observation indicates that the field dissociation does not always happen at the same location as the imaging area, and motivated our subsequent use of etched tips which we expected to have a more reproducible field emission profile. Comparing a cut PtIr tip with an etched tip, we observed a difference in the voltage needed to achieve the same field emission current at the same tip/sample separation (~ 130 nm). For a cut PtIr tip, a voltage of 50-70V was typically needed to achieve a field emission current of 0.5nA, while 90-120V was needed for etched W and PtIr tips. *A priori*, this is surprising, as the etched tips are much sharper than the cut tips on the micro-scale (< 1 micron apex diameter), and the resultant geometric field enhancement should make field emission more efficient. However, our observation of a lower voltage for cut tips indicates that cut tips are more likely to have sharp asperities that are efficient for field emission. Given this variation, we were not able to determine a threshold voltage needed for the field dissociation process.

## Atomic resolution of graphene lattice

Figure S1 shows the atomic resolution graphene imaged with molecule-terminated tip. The centers of the graphene honeycomb lattice are imaged with dark contrast.

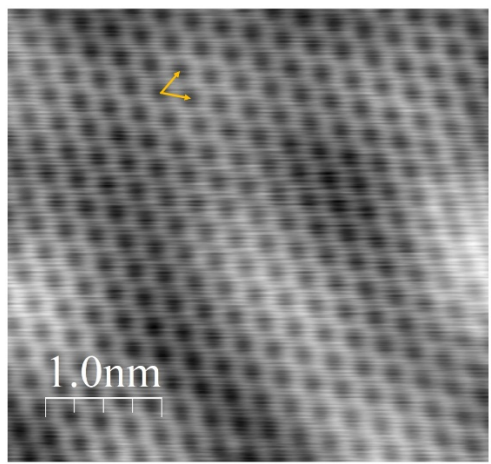

Figure S1. Atomic resolution STM image of graphene. (3nA, 50mV)

## Moiré pattern of pristine and hydrogenated graphene

Figure S2a shows an STM image of the pristine graphene island discussed in the main text (c.f., Fig. 2). Here, we see bright defects on the graphene island (marked with arrows) which were not seen in Fig 2 in the main text. Moiré pattern is also observed on graphene with lattice spacing of 6.5 nm, corresponding to a graphene lattice orientation angle of 0° relative to the underlying Cu(111) lattice. The hydrogenated graphene also has the same Moiré pattern as seen in Figure S2b, although the imaging contrast has reversed. The FFTs of both pristine and hydrogenated graphene (inset in Figure S2a and b) show the same hexagonal spots due to the Moiré pattern. Figure S2b also shows the 4.3 Å hydrogenated graphene lattice over the entire graphene island, verified in the the larger scale FFT which shows two hexagonal spot patterns (Fig. S2c).

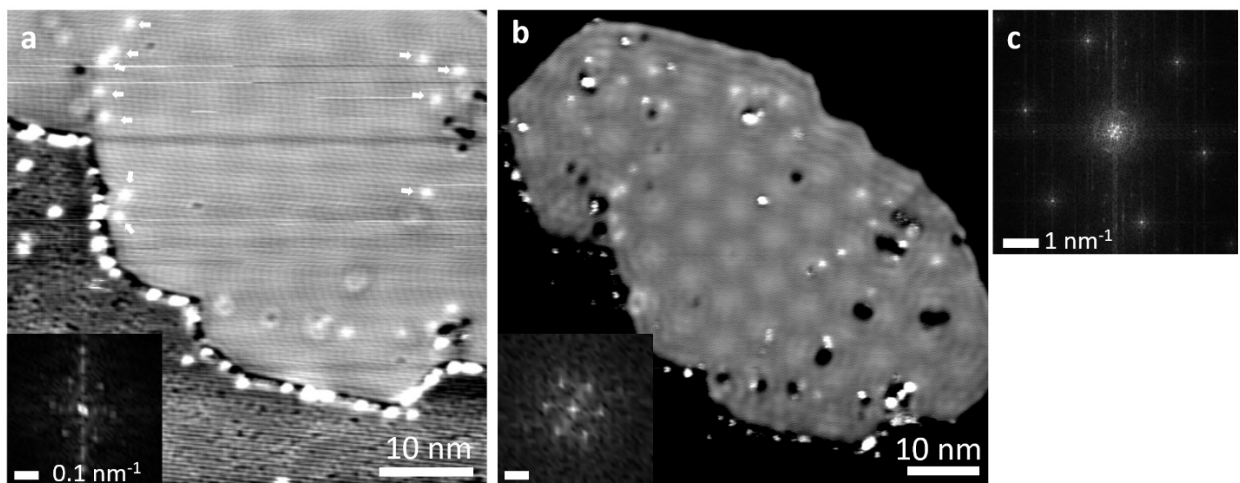

Figure S2. Moiré pattern comparison of pristine and hydrogenated graphene. (a) STM image of pristine graphene with visible Moiré pattern (appears with dark contrast) consistent with a 0 deg rotation with respect to the Cu(111) surface. Arrows show vacancy defects in graphene. The FFT inset shows hexagonal spots from the Moiré pattern. (1.0V, 0.2nA). (b) STM image of the hydrogenated graphene island with the 4.3Å lattice apparent on the entire island. The Moiré pattern (appears bright) is unchanged from the pristine graphene. (inset) The FFT shows the hexagonal spots due to the Moiré pattern, which is the same as in (a). (50mV, 0.2nA). (c) Full range FFT of the hydrogenated graphene island in (b) showing spot patterns for both the 4.3Å lattice and the Moiré pattern.

**Apparent height of point defects created after field emission dissociation**

Figure S3a shows a graphene area with new point defects generated after the field emission process. Two categories of point defects with different apparent height are observed. FigureS3b shows a line cut across the two point defects with height 15pm and 24pm.

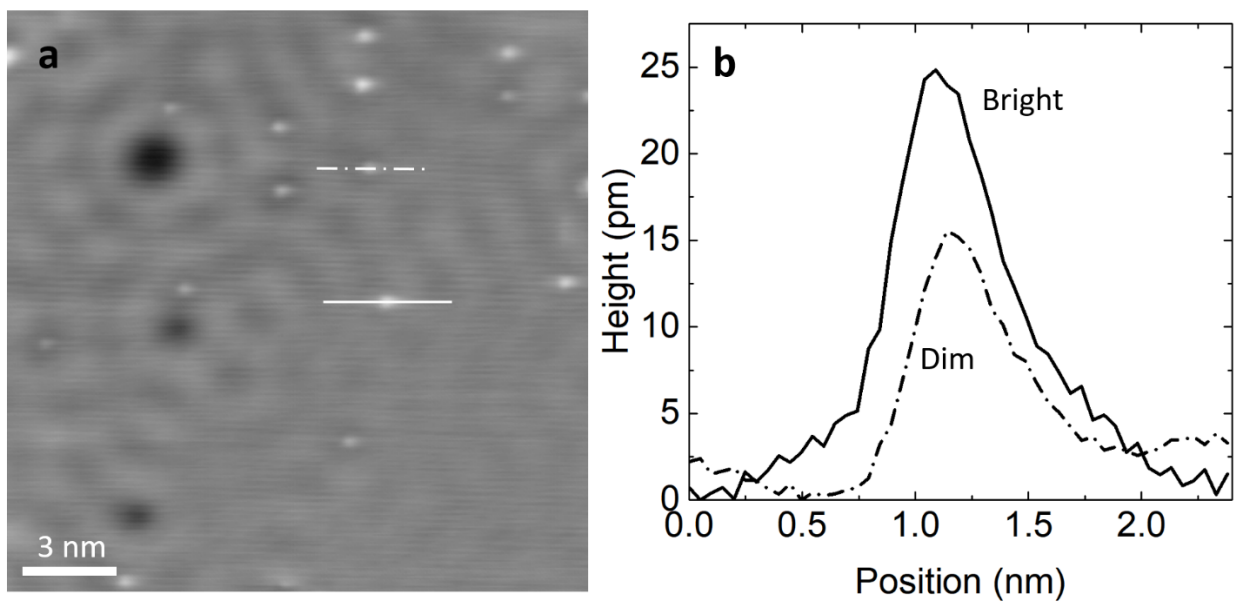

Figure S3. Apparent height of chemisorbed H following field emission. (a) STM image of graphene with chemisorbed H with two different apparent height. (50mV, 0.2nA) (b) Line cut over the dim and bright defects.

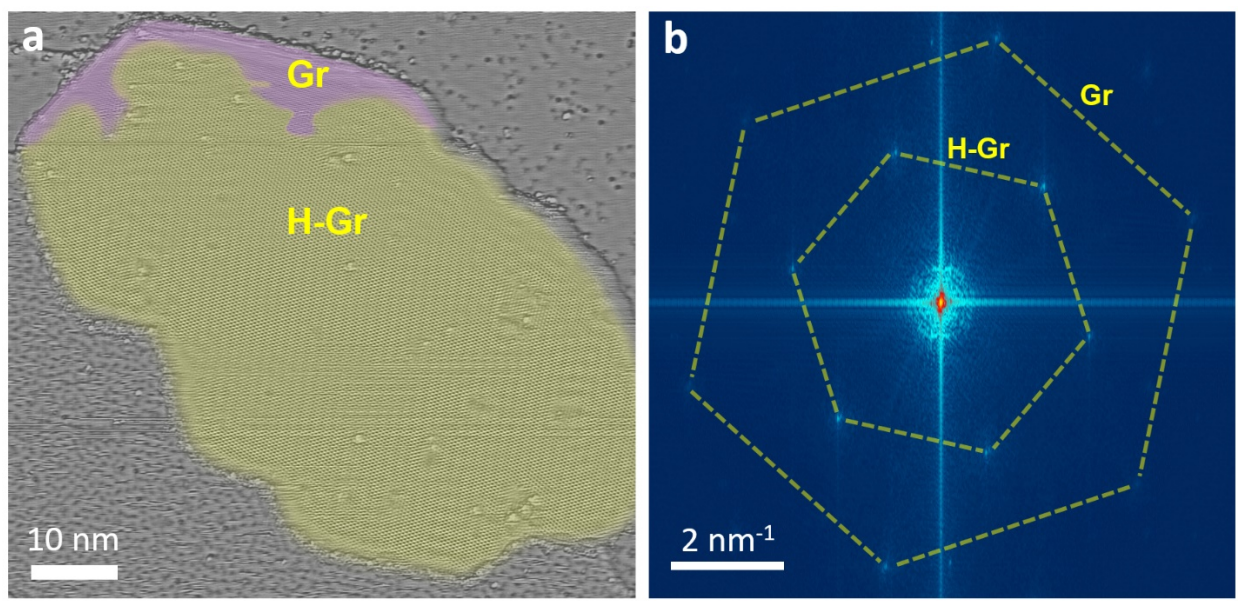

Figure S4. (a) Colorized and Laplace filtered image of the graphene island showing areas with the graphene lattice (purple) and the 4.3Å H-Gr lattice (yellow). The graphene was temporarily exposed due to perturbation by the STM tip during repeated imaging. (b) Corresponding FFT of showing spots from both the graphene and H-Gr lattices.